\documentclass[useAMS,usenatbib]{mn2e}

\usepackage{graphicx}
\usepackage{natbib}
\usepackage{aas_macros}
\begin{document}

\title[MOST satellite photometry of M67]{MOST satellite photometry of stars in the M67 field: \\ 
       Eclipsing binaries, blue stragglers and $\delta$~Scuti variables\thanks{Based 
       on photometric data from MOST, a Canadian Space Agency mission (jointly operated by
       Dynacon Inc., the University of Toronto Institute for Aerospace Studies and the University of British 
       Columbia, with the assistance of the University of Vienna), and on spectroscopic data from the 
       David Dunlap Observatory, University of Toronto}}

\author[Theodor Pribulla et al.] {Theodor Pribulla$^{1,2}$, Slavek Rucinski$^1$, 
Jaymie M. Matthews$^3$, Rainer Kuschnig$^4$, 
\newauthor
Jason F. Rowe$^3$, David B. Guenther$^5$, Anthony F. J. Moffat$^6$, Dimitar Sasselov$^7$,
\newauthor
Gordon A. H. Walker$^3$, Werner W. Weiss$^4$ \\
$^1$Department of Astronomy and Astrophysics, University of Toronto, 50 St. George Street,
Toronto, ON, M5S 3H4, Canada,\\ E-mail: pribulla@ta3.sk, rucinski@astro.utoronto.ca\\
$^2$Astronomical Institute, Slovak Academy of Sciences, 059~60 Tatransk\'a Lomnica, Slovakia\\
$^3$Department of Physics \& Astronomy, University of British Columbia, 6224 Agricultural Road, \\
Vancouver, B.C., V6T~1Z1, Canada\\
$^4$Institut f\"{u}r Astronomie, Universit\"{a}t Wien, T\"{u}rkenschanzstrasse 17, A-1180 Wien, Austria\\
$^5$Institute for Computational Astrophysics, Department of Astronomy and Physics,
Saint Mary's University, \\  Halifax, NS, B3H~3C3, Canada\\
$^6$Observatoire Astronomique du Mont M\'{e}gantic, D\'{e}partment de Physique, Universit\'{e}
de Montr\'{e}al, C.P.6128, \\  Succursale: Centre-Ville, Montr\'{e}al, QC, H3C~3J7, Canada\\
$^7$Harvard-Smithsonian Center for Astrophysics, 60 Garden Street, Cambridge, MA 02138, USA\\
}

\date{Accepted 0000 Month 00. Received 0000 Month 00; 
in original form 2007 March 17}

\pagerange{\pageref{firstpage}--\pageref{lastpage}} \pubyear{2008}

\maketitle

\label{firstpage}

\begin{abstract}
We present two series of MOST (Microvariability \& Oscillations of STars) 
space-based photometry, covering nearly continuously 10 days in 2004 and 30 days
in 2007, of selected variable stars in the upper Main Sequence of the old open
cluster M67. New high-precision light curves were obtained for the 
blue-straggler binary/triple systems AH~Cnc, ES~Cnc and EV~Cnc. The precision
and phase coverage of ES~Cnc and EV~Cnc is by far superior to any previous
observations. The light curve
of ES~Cnc is modelled in detail, assuming two dark photospheric spots and 
Roche geometry. An analysis of the light curve of AH~Cnc indicates a low mass
ratio ($q \sim 0.13$) and a high inclination angle for this system. Two new
long-period eclipsing binaries, GSC~814--323 and HD~75638 (non-members of M67)
were discovered. We also present ground-based DDO spectroscopy of ES~Cnc
and of the newly found eclipsing binaries. Especially interesting is HD~75638,
a member of a visual binary, which must itself be a triple or a
higher-multiplicity system. New light curves of two $\delta$~Scuti pulsators,
EX~Cnc and EW~Cnc, have been analyzed leading to detection of
26 and 8 pulsation frequencies of high temporal stability.
\end{abstract}


\section{Introduction}
\label{intro}

M67 (NGC 2682) is one of the best studied open clusters. It is also one of the
oldest known galactic clusters, with an estimated age of about 4 Gyr. Other
parameters of the cluster are: $E(B-V) = 0.05$ and distance modulus $(m-M_V) =
9.59$ \citep{mmj1993}. M67 contains a high percentage of blue stragglers, 
stars which are bluer and more luminous than the main sequence turnoff point of 
the cluster. For M67, the turnoff occurs near $B-V = 0.55$, and $V = 13.0$. The
formation mechanism of blue straggles is still debated, but the most probable
scenario is via mass transfer or mergers in close binaries \citep{bail1995}.
Several blue stragglers in M67 are known to be eclipsing binaries, which
supports this hypothesis. The large number of close binaries in M67 is
suggested by the relatively large number (25) of X-ray sources 
\citep{bell1998}. In such an old cluster, the X-ray sources cannot be
single rapidly-rotating stars, but -- more likely --  are tidally locked
components of close binaries. In addition, long-term spectroscopic surveys of
M67 \citep{milo1992,lath1996} revealed among the blue stragglers, 
5 spectroscopic binaries with orbital periods between 850 -- 5000 days.

In this paper, we present two time series of rapid,
quasi-continuous photometry for selected
targets in M67 obtained with the MOST (Microvariability \& Oscillations of 
STars) satellite in 2004 and 2007. The targets were selected primarily to study
two domains of cluster objects accessible to a small telescope like MOST: the
upper Main Sequence (mostly blue stragglers) and red giants. Here we describe
the results for the former group of objects, including two $\delta$~Scuti, 
while a future paper (Kallinger et al., in 
preparation) will describe analyses and asteroseismology of the p-mode
pulsations of the red giants in the cluster.

The main goals of the observations in the context of this paper were twofold: 
(i)~a search for new variables with very small amplitudes, and (ii)~obtaining
high-quality light curves, particularly for blue stragglers with periods close
to one day which are difficult to sample properly with ground-based
observations. As we describe in Section~\ref{observe}, the 2004 observations
were experimental and short (compared to other MOST runs), while the scope of
the 2007 observations was much expanded to include time series photometry of a
large number of guide stars in the M67 field. The photometric observations of
MOST have been augmented by medium-dispersion spectroscopy performed at the
David Dunlap Observatory (DDO).

The paper is organized as follows: Section~\ref{observe}
presents MOST time-series photometry
and spectroscopic observations performed at the DDO. The light variations
are described in Section~\ref{lightcurves} while a detailed discussion
of individual binary systems known or found to be variable is given in
Section~\ref{results}. The $\delta$~Scuti pulsating stars are discussed
separately in Section~\ref{deltascuti}. The paper
is concluded with the summary of the results in Section~\ref{sum}.

\section{Observations}
\label{observe}
\subsection{MOST photometry}

MOST (Microvariability \& Oscillations of STars) is a microsatellite housing
a 15-cm telescope which feeds a CCD photometer through a single, custom,
broadband optical filter (350 - 700 nm). The pre-launch characteristics of the
mission are described by \citet{WM2003} and the initial post-launch performance
by \citet{M2004}. MOST is in a Sun-synchronous polar orbit (820 km altitude) 
from which it can monitor some stars for as long as 2 months without
interruption. The instrument was designed to obtain highly precise photometry
of single, bright stars through Fabry lens imaging. Since the launch, the
satellite
capabilities have been expanded to obtain Direct Imaging photometry 
of fainter stars in sub-rasters of the open area of the Science CCD, and to
obtain photometry of the guide stars used to orient the spacecraft. Both
observing modes yield the same time sampling and coverage as the Primary
Science Target in the MOST field.

\begin{figure}
\includegraphics[width=84mm,clip=]{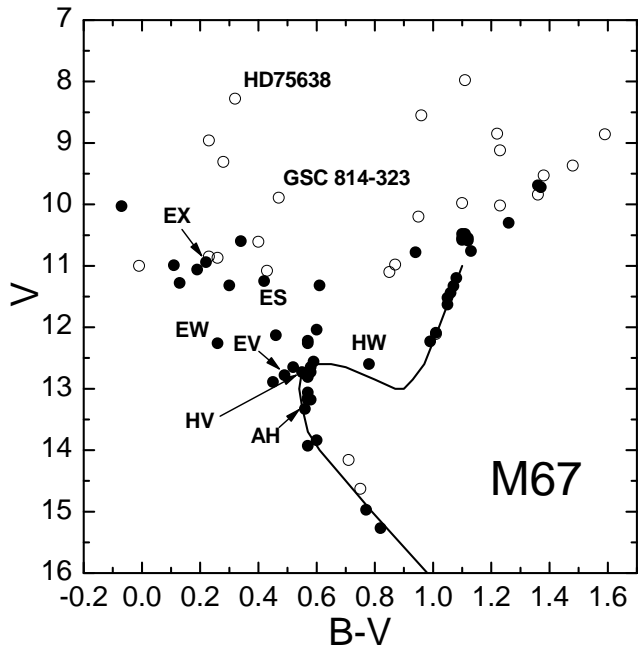}
\caption{The colour -- magnitude diagram of M67 indicating all targets observed
         by MOST in the field. The M67 members are marked by filled circles
while non-members are marked 
by open circles. The variables reported in this paper are
         labelled by their variable star names. The $(B-V)$ colours and $V$ 
         magnitudes are from \citet{mmj1993}; for non-members, we adopted the
         data from the TYCHO2 catalogue \citep{Tycho2}. The solid curve 
         represents the fiducial isochrone for M67 \label{fig01}.}
\end{figure}

MOST was never originally intended to obtain photometry of stars fainter than
$V \sim 6$, nor of multiple targets in a stellar cluster field. Its pre-launch
mission goals were asteroseismology of bright 
solar-type stars, Wolf-Rayet and magnetic
pulsating stars. The lack of multicolour measurements with MOST limits detailed
modelling of the light curves of eclipsing binaries, particularly with 
components of very different spectral types. However, the ability to monitor
such systems nearly continuously for weeks at a time leads to excellent light 
curves which are useful 
for variable star detection and for studies of variables whose periods
are close to an integer number of days.

With 3 arcsec pixels and the $1k \times 1k$ CCD detector,
the field of view of the open area of the MOST Science CCD is about 
$0.75\degr \times 0.5\degr$ (part of the field is taken by the Fabry lenses).
This is larger than the size of the inner part of the M67
cluster as defined by its brightest stars. 
As a result, we could include as guide stars and obtain photometry for 
several stars which are close to M67, but are not members of the cluster. 
Two of these stars (HD~75638 and 
GSC~814-323) were found to be eclipsing binaries (or multiple systems containing
an eclipsing binary). 

The first series of M67 observations was 10 days long, during 13 - 23
February 2004 and was designed 
as a test of MOST extended capability. In this run, 30 objects
were observed in the direct imaging mode in small sub-rasters of various sizes.
The second run covered 30 days during 12 February - 14 March 2007.  The 
pointing performance of the spacecraft had been dramatically improved by this
time, and the guide star photometric capability had been implemented. In this
run, about 40 guide stars were monitored with an 
additional 6 stars observed as direct imaging targets. About
3/4 through this second observing run (on 5 March 2007), the roll angle of the
satellite was changed to reduce the amount of scattered Earthshine on the MOST
focal plane.  As a result of the change in focal plane orientation, 38 of the 
original 42 guide stars could be retained, and 6 new ones were added for the
remainder of the run. There was no change in the selection of direct imaging 
targets after the roll change. All the MOST M67 targets are marked in the 
colour -- magnitude diagram of M67 in Fig.~\ref{fig01}. 

MOST Guide Star photometry was processed on board the satellite. Direct
imaging data were processed on the ground, after downloading of the six 
$20{\times}20$-pixel sub-rasters (for the 2007 run) centred on target stars, 
much like conventional CCD photometry. In January 2006, MOST lost use of its
startracker CCD due to a severe charged particle strike, and since then, 
both satellite attitude control and science functions have been shared by the
Science CCD. This means that individual exposures must be now relatively short (less
than 3.5 sec) to update the pointing information sufficiently rapidly. Hence,
we stack exposures on board to achieve high $S/N$ values. During the 2007 run, stacks
consisted typically of 14 exposures, each 1.5 sec long. For the faintest stars,
the cumulative CCD readout noise for the stacked exposures is the dominant
noise source.

A list of the observed stars is given in Table~\ref{tab01}. All the raw data
and the reduced light curves presented in this paper are available in the MOST
Public Data Archive, accessed through the Science Page of the MOST Mission web
site at {\sf www.astro.ubc.ca/MOST}. 

In 2007, six known blue stragglers (Sanders 1036, 1082, 1263, 1280,
1282, 1284) were observed in the direct imaging mode. S1082 (ES~Cnc) was
monitored as a guide star and in the direct imaging mode. The latter mode,
being similar to conventional CCD area photometry, produces the
more precise photometry. An attempt was made to analyse photometry of five
guide stars (HD 75784, BD +13$\degr$2006, HD 75717, SAO 98164 and GSC~816-2629)
used for the 2004 run, but the short exposures (and lack of onboard stacking)
and the larger pointing wander experienced earlier in the MOST mission, meant
a relatively large photometric scatter for those targets. The focal
plane scale of 3 arcsec per pixel, the pointing wander of a few pixels in the
2004 data and the typical FWHM of 2.5 pixels led to blending of images 
which was a serious concern for visual pairs separated 
by less than 10--15 arcsec.

\begin{table*}
\begin{scriptsize}
\caption{List of stars in M67 observed by the MOST satellite \label{tab01}}
\begin{center}
\begin{tabular}{cccrrrrlcccll}
\hline
  GSC    &  $\alpha$   &  $\delta$  &  $V$  & $B-V$ & Sand.& MMJ  & Other      &M67& Obs       & Sp. type & Type   \\
\hline                                                                                        
813-1617 & 08 49 34.65 & 11 51 25.7 &  9.53 &  1.38 &  258 & 6469 & BD+12 1913 & N & {\tt -GG} & K2       &        \\
813-2033 & 08 49 43.46 & 11 48 38.9 & 11.00 &$-0.01$&  376 &      &            & N & {\tt -GG} &          &        \\
813-2806 & 08 49 50.15 & 11 25 33.7 & 10.20 &  0.95 &  335 &      & BD+11 1920 & N & {\tt --G} &          &        \\
813-1719 & 08 49 55.22 & 11 15 10.8 &  8.55 &  0.96 &      &      & SAO98150   & ? & {\tt --G} & G5       &        \\
813-2344 & 08 49 56.83 & 11 41 33.1 &  9.84 &  1.36 &  364 & 6470 & BD+12 1917 & N & {\tt --G} &          &        \\
813-2339 & 08 50 00.59 & 11 43 58.0 & 10.98 &  0.87 &  477 &      &            & N & {\tt -GG} &          &        \\
813-1152 & 08 50 02.40 & 11 55 24.8 &  9.12 &  1.23 &  495 &      & BD+12 1918 & ? & {\tt -GG} & K0       &        \\
813-1032 & 08 50 12.30 & 11 51 24.5 &  8.86 &  1.59 &  488 & 6471 & BD+12 1919 & N & {\tt -GG} & K5       &        \\
813-1419 & 08 50 18.29 & 11 55 21.4 &  9.98 &  1.10 &  494 & 6472 & BD+12 1920 & N & {\tt --G} &          &        \\
813-2055 & 08 50 18.62 & 11 24 28.4 &  8.85 &  1.22 &  440 &      & SAO98156   & N & {\tt --G} & K0       &        \\
813-2651 & 08 50 19.91 & 11 21 30.6 &  9.31 &  0.28 &  436 &      & SAO98157   & N & {\tt --G} & A5       &        \\
816-2629 & 08 50 23.71 & 12 56 14.8 & 10.61 &  0.40 &      &      &            & N & {\tt G--} &          &        \\
816-1780 & 08 50 50.83 & 12 59 05.2 & 10.85 &  0.23 &      &      & BD+13 1999 & N & {\tt G--} & K5       &        \\  
813-1353 & 08 50 55.70 & 11 52 14.7 & 12.04 &  0.60 &  792 & 6477 &            & Y & {\tt DGG} & F5III    & BS     \\
813-2069 & 08 50 56.15 & 11 51 55.7 & 14.63 &  0.75 &  790 & 5302 &            & N & {\tt D--} & G7V      &        \\
813-2019 & 08 51 02.03 & 11 45 18.6 & 14.97 &  0.77 &  754 & 5372 &            & Y & {\tt D--} & G7V      &        \\
813-2294 & 08 51 03.47 & 11 45 01.8 & 11.32 &  0.30 &  752 & 6476 &            & Y & {\tt DGG} & A7.2:m   & BS     \\
814-1795 & 08 51 11.76 & 11 45 22.0 & 10.03 &$-0.07$&  977 & 6481 &            & Y & {\tt -GG} & B8V      & BS     \\
814-1099 & 08 51 12.68 & 11 52 42.2 & 10.59 &  1.12 & 1074 & 6492 &            & Y & {\tt -GG} & G8III    & RG     \\ 
814-1931 & 08 51 14.32 & 11 45 00.0 & 11.08 &  0.43 &  975 & 6480 &            & N & {\tt -GG} & F3       &        \\
814-1491 & 08 51 14.78 & 11 47 24.2 & 12.81 &  0.57 & 1003 & 5562 &            & Y & {\tt D--} & F8V      &        \\ 
814-1647 & 08 51 15.46 & 11 47 31.6 & 12.65 &  0.52 & 1005 & 5571 &            & Y & {\tt D--} & F2       & BS     \\
814-1735 & 08 51 16.84 & 11 45 41.8 & 14.16 &  0.71 &  981 & 5594 &            & N & {\tt D--} & G6V      &        \\
814-1763 & 08 51 16.98 & 11 50 44.6 & 11.20 &  1.08 & 1054 & 6489 &            & Y & {\tt -GG} & K0III-IV & RG,vis \\ 
814-1493 & 08 51 17.10 & 11 48 16.1 & 10.30 &  1.26 & 1016 & 6486 &            & Y & {\tt -GG} & K2III    & RG     \\ 
814-1685 & 08 51 17.36 & 11 47 00.7 & 13.06 &  0.57 &  998 & 5610 &            & Y & {\tt D--} &          &        \\
814-2439 & 08 51 17.37 & 11 46 03.3 & 13.93 &  0.57 &  987 & 5608 &            & Y & {\tt D--} & G0V      &        \\
814-2331 & 08 51 17.48 & 11 45 22.7 &  9.72 &  1.37 &  978 & 6482 &            & Y & {\tt -GG} & K4III    & RG     \\
814-1827 & 08 51 18.01 & 11 45 54.3 & 12.73 &  0.55 &  986 & 5624 & HV~Cnc     & Y & {\tt D--} &          & EB     \\
814-1803 & 08 51 18.70 & 11 47 02.9 & 12.60 &  0.78 &  999 & 5643 & HW~Cnc     & Y & {\tt D--} &          &        \\
814-1937 & 08 51 19.93 & 11 47 00.5 & 12.13 &  0.46 &  997 & 5667 &            & Y & {\tt D--} & F1       & BS     \\
814-2405 & 08 51 20.10 & 12 18 10.4 &  9.37 &  1.48 & 1135 & 6495 & BD+12~1924 & ? & {\tt -G-} & K2       &        \\
814-1997 & 08 51 20.14 & 11 46 41.9 & 12.76 &  0.56 &  995 & 5675 & NSV18058   & Y & {\tt D--} &          &        \\
814-0963 & 08 51 20.17 & 11 52 48.2 & 13.84 &  0.60 & 1075 & 5692 &            & Y & {\tt D--} &          &        \\
814-1975 & 08 51 20.35 & 11 45 52.6 & 13.15 &  0.57 & 2205 & 5679 &            & Y & {\tt D--} &          &        \\
814-1811 & 08 51 20.56 & 11 46 05.0 & 13.18 &  0.57 &  988 & 5687 &            & Y & {\tt D--} &          &        \\
814-2137 & 08 51 20.60 & 11 46 16.6 & 12.89 &  0.45 & 2204 & 5688 & NSV18059   & Y & {\tt D--} & G:       & BS     \\
814-0889 & 08 51 20.79 & 11 53 26.2 & 11.25 &  0.42 & 1082 & 6493 & ES~Cnc     & Y & {\tt DDD} & F4       & BS,EB  \\
814-1975 & 08 51 21.24 & 11 45 52.8 & 12.26 &  0.57 &  984 & 5699 &            & Y & {\tt D--} & F5       & BS     \\
814-2231 & 08 51 21.56 & 11 46 06.0 & 11.44 &  1.06 &  989 &      &            & Y & {\tt DGG} &          & RG     \\
814-1109 & 08 51 21.76 & 11 52 37.8 & 11.32 &  0.61 & 1072 & 6491 &            & Y & {\tt DGG} & G3III-IV & BS     \\
814-0793 & 08 51 21.96 & 11 53 09.1 & 15.27 &  0.82 & 1079 & 5725 &            & Y & {\tt D--} & G8V      &        \\
814-1863 & 08 51 22.07 & 11 46 41.1 & 13.18 &  0.58 &  994 & 5716 &            & Y & {\tt D--} &          &        \\
814-1531 & 08 51 22.81 & 11 48 01.7 & 10.48 &  1.11 & 1010 & 6485 &            & Y & {\tt -GG} & K2III    & RG     \\
814-1025 & 08 51 26.18 & 11 53 51.8 & 10.48 &  1.10 & 1084 & 6494 &            & Y & {\tt -GG} & G8III    & RG     \\
814-1911 & 08 51 26.44 & 11 43 50.6 & 11.28 &  0.13 &  968 & 6479 &            & Y & {\tt DGG} & A4.5:m   & BS     \\
814-1205 & 08 51 27.02 & 11 51 52.5 & 10.99 &  0.11 & 1066 & 6490 &            & Y & {\tt -GG} & A2V      & BS     \\
814-0601 & 08 51 27.44 & 12 07 40.7 &  8.28 &  0.32 &      &      & HD~75638   & N & {\tt -G-} & F0       & EB,vis \\
814-2319 & 08 51 28.15 & 11 49 27.4 & 12.78 &  0.49 & 1036 & 5833 & EV~Cnc     & Y & {\tt -DD} & F3       & BS,EB  \\
814-1147 & 08 51 28.98 & 11 50 33.0 & 10.55 &  1.12 & 1279 & 6503 &            & Y & {\tt -GG} & K1III    & RG     \\
814-1515 & 08 51 29.90 & 11 47 16.6 &  9.69 &  1.36 & 1250 & 6499 & BD+12 1926 & Y & {\tt -GG} & K3III    & RG     \\
814-2079 & 08 51 30.50 & 11 48 57.0 & 12.11 &  1.01 & 1264 & 5877 &            & N & {\tt -GG} & F8       & RG     \\
814-1003 & 08 51 32.59 & 11 50 40.3 & 12.26 &  0.26 & 1280 & 5940 & EW~Cnc     & Y & {\tt DDD} & F0I      & DS     \\
814-2087 & 08 51 32.76 & 11 48 51.8 & 11.06 &  0.19 & 1263 & 6501 &            & Y & {\tt -DD} & A7III    & BS     \\
814-1315 & 08 51 34.32 & 11 51 10.4 & 10.94 &  0.22 & 1284 & 6504 & EX~Cnc     & Y & {\tt -DD} & A7       & BS,DS  \\
814-0741 & 08 51 35.79 & 11 53 34.7 & 12.23 &  0.99 & 1305 & 5997 &            & Y & {\tt -GG} & K0IV     & RG     \\
814-0847 & 08 51 37.20 & 11 59 02.4 & 11.10 &  0.85 & 1327 & 6507 &            & N & {\tt -G-} & G8       &        \\
814-1849 & 08 51 37.43 & 11 50 05.4 & 12.56 &  0.59 & 1275 & 6018 &            & Y & {\tt D--} & F9IV     &        \\
814-2291 & 08 51 37.85 & 11 50 57.0 & 13.33 &  0.56 & 1282 & 6027 & AH~Cnc     & Y & {\tt -DD} & F6.5     & EB     \\
814-1981 & 08 51 39.26 & 11 50 04.0 & 12.22 &  0.57 & 1273 & 6047 &            & Y & {\tt D--} & F8V      & BS     \\
814-1225 & 08 51 39.38 & 11 51 45.5 & 12.09 &  1.01 & 1293 & 6050 &            & Y & {\tt -GG} &          & RG     \\
814-1007 & 08 51 42.34 & 11 50 07.5 & 11.63 &  1.05 & 1277 & 6502 &            & Y & {\tt -GG} &          & RG     \\
814-1471 & 08 51 42.37 & 11 51 23.0 & 11.33 &  1.07 & 1288 & 6505 &            & Y & {\tt -GG} & K0       & RG     \\
814-1823 & 08 51 43.53 & 11 44 26.1 & 10.76 &  1.13 & 1221 & 6497 &            & Y & {\tt -GG} & K1III    & RG     \\
814-1619 & 08 51 43.87 & 11 56 42.2 & 10.58 &  1.10 & 1316 & 6506 &            & Y & {\tt -GG} & G8       &        \\
814-1591 & 08 51 45.08 & 11 47 45.9 & 11.52 &  1.05 & 1254 & 6500 &            & Y & {\tt -GG} & K0III    & RG     \\
814-2047 & 08 51 48.64 & 11 49 15.5 & 10.87 &  0.26 & 1267 &      &            & ? & {\tt DGG} & A7V      &        \\
814-1171 & 08 51 49.12 & 11 49 43.8 & 12.73 &  0.58 & 1270 & 6166 &            & Y & {\tt D--} &          &        \\
814-0795 & 08 51 49.35 & 11 53 39.0 &  7.98 &  1.11 & 1306 & 6193 & HD75700    & N & {\tt -GG} & K0       &        \\
814-1463 & 08 51 49.97 & 11 49 31.4 & 12.65 &  0.58 & 1268 & 6177 &            & Y & {\tt D--} & F8       &        \\
814-1667 & 08 51 50.18 & 11 46 06.8 & 10.78 &  0.94 & 1237 & 6498 &            & Y & {\tt -GG} & G8III    & RG     \\
814-1011 & 08 51 56.01 & 11 51 25.9 & 10.60 &  0.34 & 1466 & 6511 &            & Y & {\tt -GG} & A3       & BS     \\
814-0817 & 08 51 59.51 & 11 55 04.5 & 10.55 &  1.10 & 1479 & 6512 &            & Y & {\tt -GG} & K1       & RG     \\ 
817-1563 & 08 52 01.93 & 13 23 26.3 &  8.96 &  0.23 &      &      & HD75717    & N & {\tt G--} & F2       &        \\  
814-0133 & 08 52 02.62 & 12 25 54.1 & 10.02 &  1.23 & 1533 & 6513 & BD+12 1928 & ? & {\tt -G-} & K0       &        \\
814-0323 & 08 52 10.65 & 12 17 51.9 &  9.89 &  0.47 & 1522 &      & BD+12 1929 & ? & {\tt -G-} &          & EB,vis \\
\hline                                                                                        
\end{tabular}                                                                                 
\end{center}                                                                                  
\end{scriptsize}                                                                              
\end{table*}

\begin{table*}
\addtocounter{table}{-1}
\begin{scriptsize}
\caption{(continued)}
\begin{center}
\begin{tabular}{cccrrrrlcccll}
\hline
  GSC    &  $\alpha$   &  $\delta$  &  $V$  & $B-V$ & Sand.& MMJ  & Other      &M67& Obs       & Sp. type & Type   \\
\hline                                                                                        
814-1881 & 08 52 10.72 & 11 44 05.1 & 10.70 &  0.11 & 1434 & 6510 & BD+12 1930 & ? & {\tt -GG} & A3III    &        \\
814-2361 & 08 52 18.56 & 11 44 26.2 & 10.47 &  1.12 & 1592 & 6516 &            & Y & {\tt -GG} & K0III    & RG     \\ 
817-1748 & 08 52 23.94 & 13 14 00.4 &  9.15 &  1.00 &      &      & HD75784    & N & {\tt G--} & G5       &        \\                                          
817-1469 & 08 52 24.12 & 13 16 06.2 &  9.86 &  0.39 &      &      & BD+13 2006 & N & {\tt G--} &          &        \\                                          
\hline                                                                                        
\end{tabular}                                                                                 
\end{center}                                                                                  
\end{scriptsize}                                                                              
\flushleft{Explanation of columns: GSC -- Guide Star Catalog number; $\alpha$,$\delta$ -- ICRS 
equatorial coordinates for epoch and equinox 2000; $V$, $B-V$ -- visual magnitude and colour 
index from \citet{mmj1993}; Sand.\ -- identification in \citet{sand1977}; MMJ -- identification 
of star in \citet{mmj1993}; Other -- other designation; M67 - cluster membership according to 
\citet{bala2007}; Obs -- type of MOST observations in 2004, 2007 part A, 2007 part B: ``D'' 
Direct Imaging mode, ``G''' Guide Star mode, ``-'' not observed, (ES~Cnc was observed both in 
Direct Imaging and Guide Star modes in 2007); Sp.\ type -- spectral type; Type -- general type 
and classification of the system: BS -- blue straggler \citep{bala2007}, RG -- red giant
\citep{stello2007}, EB -- eclipsing binary, vis -- visual pair, DS -- $\delta$~Scuti variable.}
\end{table*}

\subsection{DDO spectroscopy}
\label{ddo}

\begin{figure}
\includegraphics[width=84mm,clip=]{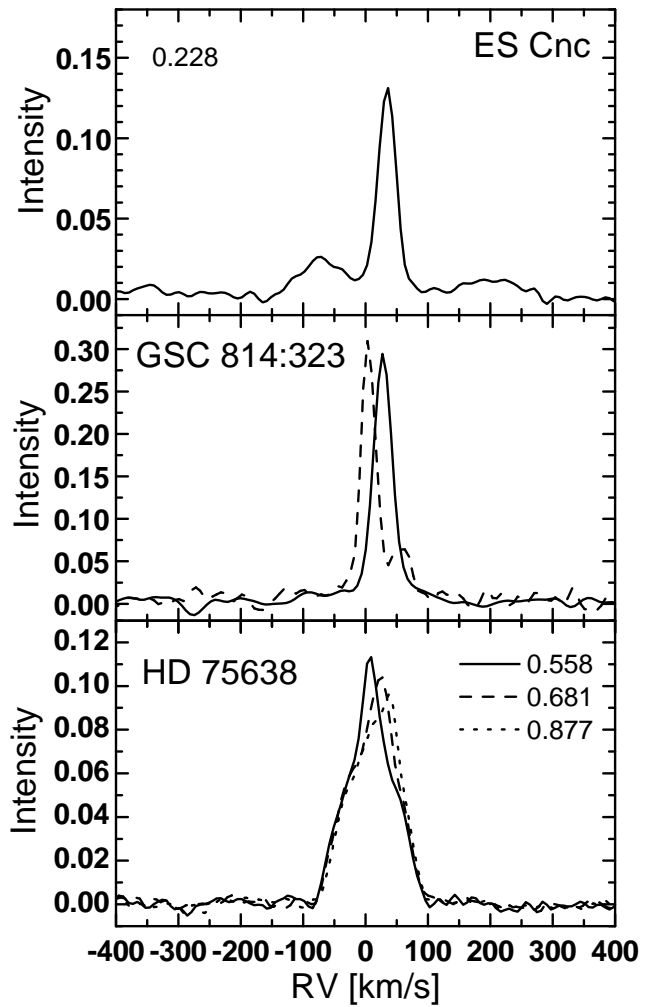}
\caption{Broadening functions (BFs) of eclipsing variables ES~Cnc, GSC~814-323
         and HD~75638. For HD~75638, we show the BFs for a few orbital
         phases calculated from the known ephemerides as in Table~\ref{tab02}. ES~Cnc
         exhibits very well defined signatures of two rapidly rotating
         components and of a third, slowly rotating component. This system 
         deserves to be the subject of further study using the BF approach.
\label{fig02}}
\end{figure}

After the two MOST photometric runs, spectroscopic observations of selected M67
targets were obtained using the slit spectrograph at the Cassegrain focus of 
the 1.88-m telescope of the David Dunlap Observatory. The spectra were taken in
a wavelength window about 240~\AA~ wide around the MgI triplet (5167, 5173 and 
5184~\AA) with an effective resolving power of $R = 12,000 - 14,000$. The 
diffraction grating with 2160 lines/mm was used giving the scale of 
0.117~\AA/pixel. One-dimensional spectra were extracted by the usual procedures
within the IRAF environment \footnote{IRAF is distributed by the National
Optical Astronomy Observatories, which are operated by the Association of
Universities for Research in Astronomy, Inc., under cooperative agreement with
the NSF.} after bias subtraction and flat field division. Cosmic ray trails were
removed using the program of \citet{pych2004}. The observations were analysed
with the technique of broadening functions (BFs), described by \citet{Rci1992} 
and \citet{Rci2002}, using spectra of slowly rotating stars of similar spectral
type as templates. A sample of extracted BFs is given in Fig.~\ref{fig02}.

The DDO observations are part of a more extensive spectroscopic survey
(Pribulla et al., 2008, in preparation)
of variable stars detected or observed during the MOST mission, especially 
those for which no reliable spectral classifications exist in the 
literature. The goals of this survey are to:
(i)~determine the spectral type and luminosity class, (ii)~help identify the
type of variability in conjunction with the MOST photometry, (iii)~measure 
the projected rotational velocity $v \sin i$, and to (iv)~detect and study 
the spectroscopic binaries and multiple systems in the sample (for the
detailed discussion of spectroscopic observations and results see Pribulla
et al., 2008, to be submitted to MNRAS).

\section{Light curves}
\label{lightcurves}

\begin{figure}
\includegraphics[width=84mm,clip=]{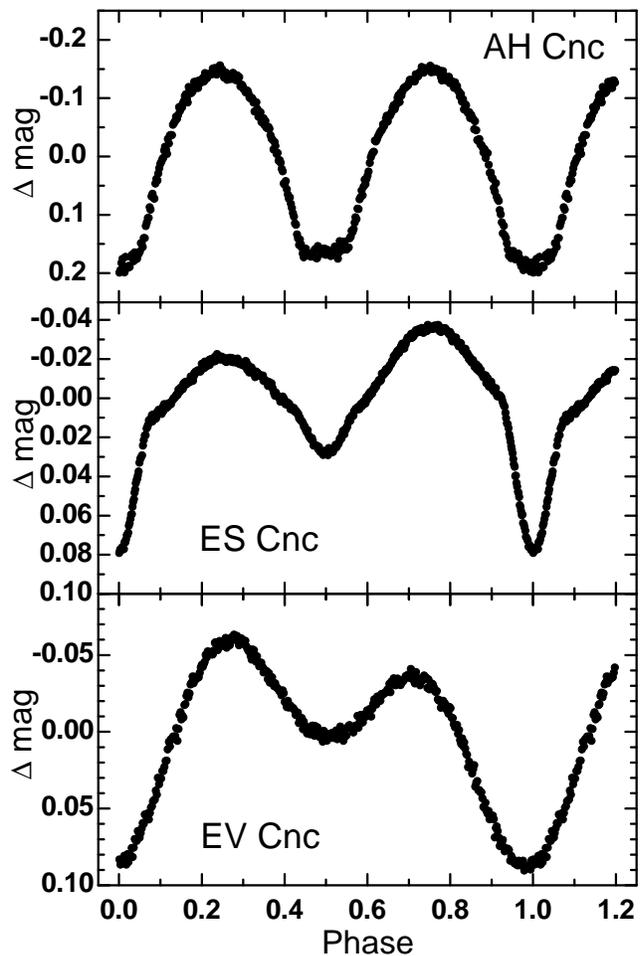}
\caption{The phased light curves of three short-period eclipsing variable 
         stars. AH~Cnc and EV~Cnc are contact binaries; ES~Cnc is most probably
         a triple system containing an eclipsing contact binary. The light 
         curves shown here are based on average data points computed from the 
         entire MOST 2007 observing run. Note the very different vertical scales
used in each of the panels.
\label{fig03}}
\end{figure}

For the known short-period eclipsing systems (AH~Cnc, ES~Cnc and EV~Cnc) we 
constructed the phased light curves shown in Fig.~\ref{fig03}. The data were
first phased using linear ephemerides, determined using both published and new
times of minimum light from the MOST photometry. Then the data were 
3-$\sigma$-clipped to remove obvious outliers resulting mostly from cosmic ray
hits. The number of outliers is strongly correlated with passages of the 
satellite through the SAA (South Atlantic Anomaly). Our final adopted 
parameters for AH~Cnc, ES~Cnc and EV~Cnc are listed in Table~\ref{tab02}. 

\begin{figure}
\includegraphics[width=84mm,clip=]{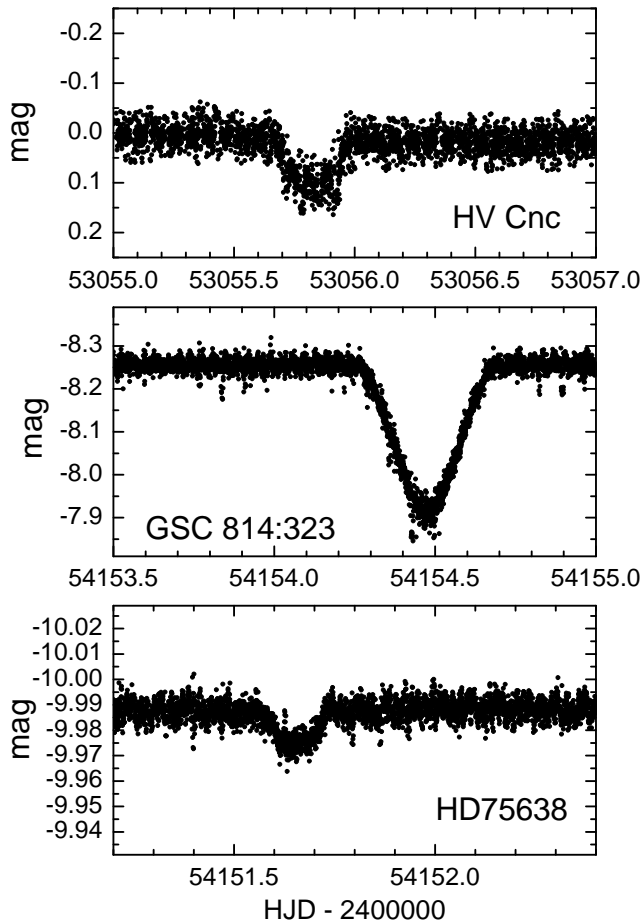}
\caption{The light curves of the long-period eclipsing variables in the MOST 
         M67 data. GSC~814--323 and HD~75638 are systems discovered during
         the MOST 2007 observing run while HV~Cnc is an already known eclipsing
         system.
\label{fig04}}
\end{figure}

In addition to the short-period binaries, we observed the long-period
eclipsing binary HV~Cnc and discovered two new eclipsing systems 
of this type, GSC~814--323 and HD75638. Light curves for phases close to
eclipses for these three systems are plotted in Fig.~\ref{fig04}. Our new
minima for the systems, determined using the \citet{kw1956} method, are listed 
in Table~\ref{tab03}.

\begin{figure}
\includegraphics[width=84mm,clip=]{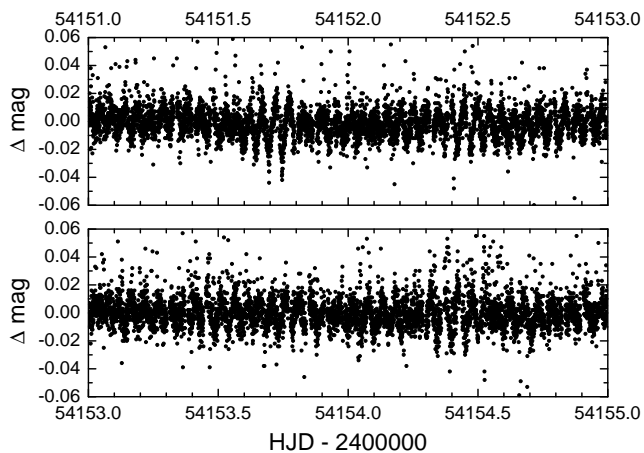}
\caption{A four-day segment of the 2007 light curves of the $\delta$~Scuti variable 
         EX~Cnc. The wavelike variations correspond to a superposition of several
         close periodicities.
\label{fig05}}
\end{figure}

MOST photometry includes two known $\delta$~Scuti variables, EX~Cnc and EW~Cnc, and
several blue stragglers which could possibly show period or quasi-periodic variations.
The pulsations of EX~Cnc are obvious and show beats of the two close periods 
(Fig.~\ref{fig05}). We discuss photometry of the $\delta$~Scuti variables in
Section~\ref{deltascuti}.

\begin{table}
\begin{scriptsize}
\caption{Parameters of the eclipsing systems AH~Cnc, ES~Cnc, and EV~Cnc
         \label{tab02}}
\begin{center}
\begin{tabular}{ccccc}
\hline
                 &   AH~Cnc      &  \multicolumn{2}{c}{ES~Cnc}        &   EV~Cnc      \\
Year              &   2007        & 2004           & 2007              &    2007       \\
$T_0$             & 52790.9042(3) & \multicolumn{2}{c}{52254.7811(11)} & 53245.087(4)  \\
Period            & 0.36045713(21)& \multicolumn{2}{c}{1.0677968(4)}   & 0.4414399(17) \\
$\Delta V_p$      &   0.33        &  0.114         &  0.121            &  0.148        \\
$\Delta V_s$      &   0.31        &  0.064         &  0.079            &  0.063        \\
Max$_1 -$ Max$_2$ &   0.00        &  0.016         &  0.027            & $-0.023$      \\
\hline
\end{tabular}
\end{center}
\end{scriptsize}
\flushleft{\footnotesize Explanation of columns: $T_0$, Period -- the linear ephemeris determined 
           from our data as well as all published minima times, $\Delta V_p$ and $\Delta V_s$ 
           -- amplitudes of the primary and secondary minima, Max$_1 - $Max$_2$ 
           -- the O'Connell effect (positive if maximum following the primary minimum
           is fainter of the two).}
\end{table}

\section{Eclipsing systems and blue stragglers}
\label{results}

\subsection{AH~Cnc}

AH~Cnc (S1282, MMJ~6027) is a contact binary which is a member of the M67 
cluster. The variability of this rather faint system ($V_{max} = 13.31$) was
detected by \citet{kura1960}. Later the system was thoroughly studied by 
\citet{whel1979}, who obtained both spectroscopic and photometric observations. 
The authors estimated its spectral type as F7V. The radial velocity orbit was,
however, of limited quality. The geometric parameters were estimated as 
$i \sim 65-68 \degr$, $K_1 \sim 100$ km~s$^{-1}$, with a very uncertain mass
ratio $q \sim 0.42 - 0.75$. An inspection of their light curves, with amplitude
of about 0.38 mag, suggested partial eclipses and therefore a relatively high
mass ratio. 

Light curves of higher precision obtained by \cite{gill1991} gave a very
different picture: The system definitely showed total eclipses. The shallower of
the two was flat, hence AH~Cnc must be an A-type contact binary with the more 
massive component being hotter. \citet{sand2003a} presented a photometric 
analysis of their light curve indicating that the mass ratio can be as low as 
0.16. More recent light curves from \citet{zhan2005} support a low mass ratio 
and clearly show an asymmetry which was interpreted as the presence of 
a dark photospheric spot. Finally, \citet{qian2006} reported photometric 
evidence of a faint third component, contributing at most only about 
$L_3/(L_1+L_2+L_3) \approx 0.01$ and causing a cyclic ($P = 36.5$ yr) variation in 
the observed times of minima.

Our phased light curve of AH~Cnc is plotted in Fig.~\ref{fig03}. The secondary 
minimum is flat and the light curve does not show any evidence of asymmetries. 
We fitted the light curve with the contact geometry
Roche model, with the bolometric albedo and 
gravity darkening coefficients appropriate for the convective envelope. The
effective temperature of the primary was set at $T_1 = 6220 K$, corresponding to 
the F7V spectral type (and supported by its colour, see Table~\ref{tab01}). 
Due to the strong correlation between the third light and mass ratio in the 
solution, for simplicity, we adopted zero third light. The resulting 
photometric elements, given in Table~\ref{tab04}, confirm the low mass ratio,
with the orbit oriented to us practically edge-on. Checks on the
consistency of the solution can be made
using the absolute magnitude calibration of \citet{rd1997}. Using the measured
orbital period (Table~\ref{tab02}) and de-reddened colour $(B-V)_0 = 0.51$, we
obtain $M_V = 3.78$ with $V_{max} = 13.15$ (assuming $A_V = 3.2 
E_{B-V}$). This gives the distance modulus of $(m-M_V) = 9.37\pm$0.02\footnote{uncertainty
is here mainly given by intrinsic differences between the individual binary systems 
and can be estimated as $\pm$0.2 mag}, close to 
the value found for M67 by \citet{niss1987}: $(m-M_V) = 9.61\pm0.04$. 
This consistent solution argues strongly against any third light in the system; 
such a third light would mean a lower brightness of the contact 
binary itself and a higher distance modulus, placing the star behind the cluster. 


\begin{figure}
\includegraphics[width=84mm,clip=]{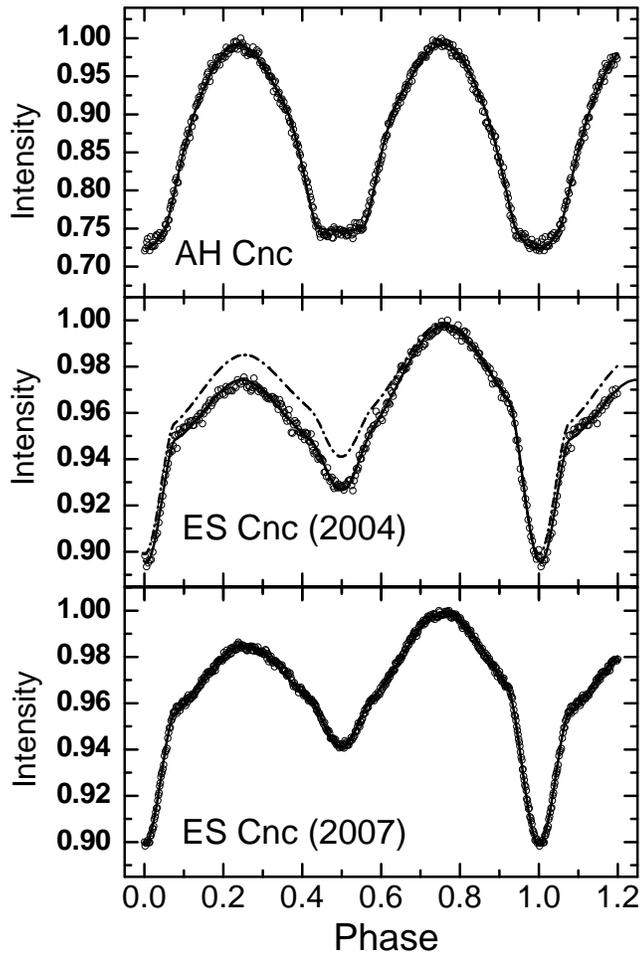}
\caption{The best fits to the light curves of AH~Cnc (top) and both MOST 2004
         and 2007 light curves of ES~Cnc obtained assuming the contact model. In 
    the     case of ES~Cnc, two cool spots on the primary component are necessary
         to explain the observed asymmetry. The panel with the 2004 light curve 
         contains for comparison the shape of the 2007 light curve 
(dot-dashed line). 
\label{fig06}}
\end{figure}

\begin{table}
\begin{scriptsize}
\caption{New primary minima of eclipsing systems determined from the MOST 
         observations. The epochs for ES~Cnc and EV~Cnc correspond to 
         ephemerides given in Table~\ref{tab02}. The epoch for the minimum of
HV~Cnc is counted from $T_0$ in the ephemeris of \citet{sand2003b}. \label{tab03}
         }
\begin{center}
\begin{tabular}{rl|rl}
\hline
 Epoch   & HJD          & Epoch     & HJD         \\
         & 2\,400\,000+ &           & 2\,400\,000+\\
\hline
\multicolumn{2}{l}{ES~Cnc} & 8918 & 54165.8639 \\
  7874 & 53051.0877   &      8920 & 54167.9957 \\
  7876 & 53053.2117   &      8922 & 54170.1339 \\
  7877 & 53054.2879   &      8923 & 54171.2019 \\
  7878 & 53055.3548   &      8924 & 54172.2696 \\
  7879 & 53056.4249   &      8925 & 54173.3370 \\
  7880 & 53057.4922   & \multicolumn{2}{l}{EV~Cnc} \\
  7881 & 53058.5610   &      8255 & 54144.1280 \\
  8898 & 54144.5101   &      8262 & 54147.2024 \\
  8899 & 54145.5763   &      8263 & 54147.6584 \\
  8901 & 54147.7092   &      8269 & 54150.3054 \\
  8902 & 54148.7871   &      8272 & 54151.6301 \\
  8903 & 54149.8446   &      8273 & 54152.0721 \\
  8904 & 54150.9149   &      8274 & 54152.5075 \\
  8905 & 54151.9802   & \multicolumn{2}{l}{HV~Cnc} \\
  8906 & 54153.0473   &       703 & 53055.828  \\
  8907 & 54154.1151   & \multicolumn{2}{l}{GSC~814-323} \\
  8908 & 54155.1824   &         0 & 54154.481  \\
  8910 & 54157.3186   & \multicolumn{2}{l}{HD~75638} \\
  8912 & 54159.4546   &         0 & 54145.85  \\
  8913 & 54160.5220   &         1 & 54151.65  \\
  8914 & 54161.5918   &         2 & 54157.47  \\
\hline
\end{tabular}
\end{center}
\end{scriptsize}
\end{table}

\subsection{ES~Cnc}

ES~Cnc (S1082, MMJ~6493, spectral type F4) is a very complex, probably best studied
blue-straggler eclipsing system in M67. Its photometric variability was 
discovered by \citet{simo1991}. Later, \citet{gore1992} discovered that it is a
close binary showing partial eclipses with an orbital period of 1.0677978(50) d.
However, \citet{milo1992} found some radial velocity variability, but no sign
of the 1.07 d period.  

A real breakthrough in the understanding of ES~Cnc 
system was made by \citet{vberg2001}, who found that in addition to the slowly 
moving, narrow-line component (later found to be moving in a 1189-d orbit 
\citep{sandal2003}), the spectrum of the system contains two other components. 
Their radial velocities vary with the orbital period of 1.07 d, so 
the system is a hierarchical triple in which all three components 
are blue stragglers. Both \citet{vberg2001} and \citet{gore1992} noted that the
maximum following the primary minimum is brighter by 0.01 -- 0.02 mag.  This, 
together with strong X-ray flux \citep{bell1998} indicates that the close pair is a 
magnetically active, RS~CVn-like eclipsing binary. Further extensive photometric 
observations by \citet{sandal2003} showed that the light curve of the system is 
variable at a level of about 0.02 mag and on a timescale of about a month. 
The authors noted a short timescale variation of the brightness, which they
interpreted as $\delta$~Sct-type variability of the third component. 

The system was comfortably bright for MOST photometry, with a typical standard 
error per individual measurement of $\sigma \sim 0.006 - 0.01$ mag. 
However, there exist some very long-term trends in the MOST 2004 and 2007 data, 
probably associated with evolution of scattered Earthshine in the field during 
the runs, so we conservatively choose not to use MOST photometry to interpret 
any slow variability in ES~Cnc. Instead, we constructed 
the phased, seasonal light curves (Fig.~\ref{fig03}) 
from $\sigma$-clipped data, to enhance the precision of our photometry to the
level of the point-to-point error of only 1.1 mmag for the 2007 data.
Our light curves show the O'Connell effect during both seasons, in 
2004 and in 2007, in contrast 
to what was reported by \citet{vberg2001} and \citet{gore1992}. In spite of
the continuous variations due to the spots and the ellipticity effect, the outer 
contacts of the eclipses are exceptionally well defined, especially in the 
2007 light curve.

The light curve of ES~Cnc was modelled first for the 2007 data.
We assumed a spectroscopic mass ratio of
$q_{sp} = 0.63$ \citep{sandal2003} and effective temperature of the primary 
$T_{eff} = 7325~K$ (the primary temperature from Table~5 of 
\citet{sandal2003}). Both the bolometric albedo and the gravity darkening 
coefficients were set following \citet{vberg2001}. The asymmetry of the light 
curve, which is very probably caused by photospheric spots, is significant but 
very stable, so we assume -- unlike \citet{vberg2001} -- synchronous
rotation of the components. The third light was optimised in the solution. 
Immediately after a few trial runs, we realized that the deformation of the 
observed light curve cannot be explained by one dark spot, and that at least 
two spots are necessary. Experimentation with the temperature factor $k$ of 
the spot (the ratio of the spot temperature and photospheric temperature 
outside the spot; \citet{linn1993}) showed that spots are rather cool with 
$k \sim 0.85$ and are small in size. Our solution (Fig.~\ref{fig06} and 
Table~\ref{tab04}) fits the observed light curve well. Because of the ill posed 
nature of this type of fitting, the solution is not unique: A fit of a similar
quality can be obtained by placing the spots at different latitudes, because 
the spot diameter correlates strongly with the temperature factor. Also, since 
the spots are not eclipsed, it is hard to determine from the photometry alone
which of the components is spotted.

The 2004 light curve was fitted with the same assumptions as for the
2007 light curve, i.e., the geometric
elements were fixed by the solution of the 2007 light curve. The 2004 phased
light curve shows a larger asymmetry than in 2007, but of the same sense: 
The maximum following the primary minimum is fainter. The resulting fit is 
shown in Fig.~\ref{fig06}. As expected, the spot radii are larger than in the
2007 fit. The spots, however, seem to have hardly moved between 2004 and 2007.

In spite of the very good reproduction of the observations by our model,  
we still regard the resulting parameters (Table~\ref{tab04}) as preliminary 
because of the large asymmetry and our inability to 
determine parameters free of spot effects (see \citet{sandal2003}). Some 
systematic uncertainty might arise from the very broad transmission of the 
MOST broadband filter and slightly variable colours
during the orbital revolution. 
The ES~Cnc system calls for thorough modelling of 
simultaneously observed line profiles (or BFs) and photometric data.

The $O-C$ diagram for all available minima of ES~Cnc shows a rather high scatter.
An attempt to interpret the deviations as due to the light-time effect 
(hereafter LITE) caused by the third component led to a substantially longer 
orbital period ($P 1304\pm13$ days) than indicated by spectroscopy, 
$P = 1188.5\pm6.76$ days \citep{sandal2003}. The problem is very probably caused by 
(i)~the low amplitude of LITE due to the short orbital period, (ii)~the shallow
eclipses making their positions rather poorly determined, and 
(iii)~a large and variable O'Connell effect causing the apparent instants 
of minima to deviate from the spectroscopic conjunctions as defined by 
the component's centres of mass. Without more highly precise and reliable 
timings of the
minima, any conclusions about the possible LITE orbit are a premature 
speculation. On the other hand, the presence of LITE would conclusively settle
the physical bond of the triple system as the third body may be simply an 
optical projection within the cluster.

\begin{table}
\begin{scriptsize}
\caption{Photometric light curve solutions for the short-period eclipsing systems 
         \label{tab04}}
\begin{center}
\begin{tabular}{lccc}
\hline
Parameter      &   AH~Cnc   &   ES~Cnc   &   ES~Cnc   \\
               &   (2007)   &   (2004)   &   (2007)   \\
\hline
$i$ [deg]      &   89.5     &     66.2   &     66.2   \\
$\Omega_1$     &   2.023    &  3.532$^f$ &  3.532     \\
$\Omega_2$     &   2.023    &  3.922$^f$ &  3.922     \\
$q$            &   0.130    &   0.63$^f$ &   0.63$^f$ \\
$T_1$ [K]      &   6300     &   7325$^f$ &   7325$^f$ \\
$T_2$ [K]      &   6368     &    5543    &    5543    \\
$l_3$          &   0.00$^f$ &    0.68    &    0.68    \\
$A_1$          &   0.50$^f$ &   1.00$^f$ &   1.00$^f$ \\
$A_2$          &   0.50$^f$ &   0.50$^f$ &   0.50$^f$ \\
$g_1$          &   0.25$^f$ &   1.00$^f$ &   1.00$^f$ \\
$g_2$          &   0.25$^f$ &   0.25$^f$ &   0.25$^f$ \\
\hline
$k^s$          &     --     &   0.80$^f$ &   0.80$^f$ \\
$r_I^s$ [deg]  &     --     &   15.35    &   11.5     \\
$l_I^s$ [deg]  &     --     &  187.2     &  179.9     \\
\hline
$r_{II}^s$ [deg] &    --    &   15.75    &   12.5     \\
$l_{II}^s$ [deg] &    --    &  293.9     &  296.6     \\
\hline
\end{tabular}
\end{center}
\end{scriptsize}
\flushleft{Explanation of rows: $i$ -- inclination angle; $\Omega$ -- 
dimensionless surface potential; $q = m_2/m_1$ -- mass ratio; $T_{1,2}$ -- 
polar temperature; $l_3 = L_3/(L_1+L_2+L_3)$ -- third light; $A_{1,2}$  -- 
bolometric albedo; $g_{1,2}$ -- gravity darkening coefficient; $k^s$ -- 
temperature factor of spot(s),  and $r_{1,2}^s$ and $l_{1,2}^s$ -- longitude 
and radius of the first and the second spot. Indices $1,2,3$ denote components 
while $I, II$ identify the spots.
}
\end{table}

\subsection{EV~Cnc}

EV~Cnc (S1036, MMJ~5833, spectral type F3) is another blue straggler binary. 
The system is very probably close to contact or perhaps already in contact, 
and is seen at low inclination. The light curve of the system is rather unusual: 
Although the light curve is continuous, indicating the contact binary nature of 
the system, the depths of minima are significantly different, hardly consistent 
with the components being in contact. Also, the maximum following the primary 
minimum is about 0.025 mag brighter than the other maximum. The system was 
initially detected by \citet{gill1991} and extensively studied 
by \citet{sand2003a}, who improved the ephemeris for 
minimum light to $T_0 = 2\,450\,500.047 + 0.441437(3) \times E$. Their 
grid-search approach to find the most probable geometric and spot parameters
led to a fairly uncertain result, exacerbated by the lack of a spectroscopic mass 
ratio, $q_{sp}$. If the system is in contact, its inclination angle is 
$30\degr < i < 38\degr$, its mass ratio is $q_{ph} \sim 0.5$ and at least two 
dark spots are required to reproduce the light curve. \citet{sand2003a} noted 
small variations in the shape of the phase diagram on timescales of about a 
month.

Our MOST photometry allowed us to determine reliably 10 new times of minimum
for EV~Cnc (Table~\ref{tab03}). Combining our new minima with the published 
ones leads to an improvement of the ephemeris to $T_0 = 2\,453\,245.087(4) 
+ 0.4414399(17) \times E$. Our light curve (Fig.~\ref{fig03}), shows a very
similar shape to that of \citet{sand2003a} or \citet{gill1991}, indicating 
very stable surface structures, similar to the behaviour found in the contact 
binary AG~Vir \citep{bell1990}. We attempted to model the light curve of EV~Cnc but we 
found that there are many combinations of parameters leading to similar
quality fits. Thus, we could add little to the discussion of \citet{sand2003a}.
Furthermore, we found that the temperature of the secondary component $T_2$ 
cannot be determined reliably because of the strong correlation between 
$T_2$, the fill-out factor, and the inclination angle. Any further progress in 
understanding EV~Cnc requires a dedicated spectroscopic study providing a 
reliable mass ratio.  

\subsection{HV~Cnc}

HV~Cnc (S986, MMJ~5624) is very probably a member of M67 \citep{bala2007}. 
Spectroscopic observations by \citet{math1990} show that the system is a 
single-lined spectroscopic binary (SB1) with a circular orbit and a period of
10.3386 days. Eclipses were observed later by \citet{sand2003b}. The
authors found another component in the spectra which they identified with a 
third star in the system, contributing about 11.5\% to the total light. 
Through a deconvolution of the photometry of all three stars, it was 
found that the primary is hotter than the turn-off point of the cluster
so it is a Blue Straggler.

HV~Cnc was observed by the MOST satellite only during the 2004 observing run. 
One primary minimum at HJD 2\,453\,055.828 was observed.  The minimum occurred 
very close to the time predicted by the ephemeris of \citet{sand2003b}, at phase 
0.99928, proving that the orbital period is very stable. No mass transfer or 
stronger interactions of components is possible for such a relatively long 
orbital period.  HV~Cnc is rather faint for MOST photometry ($V_{max}$ = 12.73), 
especially early in the mission, and the data show a scatter as large as 0.02 mag,
so that secondary minima could not be detected. No spectra of HV~Cnc were 
obtained at DDO because the system is too faint for the 1.88m telescope.


\subsection{GSC~814-323}

GSC~814-323 (S1522, BD +12$\degr$1929) is most probably not a member of M67 
\citep{zhao1993}. It is, however, a visual pair (WDS 08522+1217) consisting of
stars with $V = 9.12$ and $V = 10.21$ separated by $\rho$ = 14.4 arcsec at a 
position angle $\theta$ = 140$\degr$.

The MOST photometry revealed a single, well-defined, partial eclipse centred at 
HJD 2\,454\,154.481 and lasting about 9 hours. The brightness of the system 
dropped by 0.33 mag (Fig~\ref{fig04}). Since the system was observed in the
guide star mode (processed on board the satellite) and the MOST focal plane 
scale is 3 arcsec/pixel, the photometry is expected to include both 
visual components. 
If the brighter component is the eclipsing binary, then the corrected amplitude
of the eclipse would be about 0.48 mag. The fainter component cannot be 
responsible for the eclipse because even its total exclusion would result in 
a drop of 0.34 mag in brightness.

Unfortunately, only one eclipse was recorded during the first part of the MOST 
2007 run, setting a lower limit on the orbital period of about 11 days. The 
system was not detected as a variable during the 
ASAS\footnote{(http://archive.princeton.edu/~asas/)} nor the 
NSVS\footnote{http://skydot.lanl.gov/nsvs/nsvs.php)} variability surveys.

There is a single radial velocity measurement of GSC~814-323 at 
HJD 2\,441\,321.88 published by \citet{math1986}, giving $RV = 12.6(5)$ 
km~s$^{-1}$. The measurement probably refers to the brighter component of the 
visual pair. Two spectra taken at DDO suggest that the system
is a double-lined spectroscopic binary (SB2); see Fig. \ref{fig02}. The ratio
of intensities, as estimated from the BFs when the components are split, is 
about $L_2/L_1 \sim 0.29$. Therefore, we could expect secondary eclipses about
0.09 mag deep. More spectra are needed to find the orbital period and to
characterise this triple system.

\subsection{HD~75638}

HD75638 (GSC~814-601, BD +12$\degr$1925; spectral type F0) is another field 
star close to M67. The star is, in fact, a known visual triple system 
(WDS 08515+1208) consisting of components A~($V = 8.28$), B~($V = 10.17$) and 
C~($V = 11.8$). While the pair AB is rather tight ($\rho$ = 1.4 arcsec), the 
faintest component C is separated by 15.1 arcsec from the brightest one;
it is too far to contribute to the aperture photometry used for the
MOST data. 

HD~75638 (components AB) was found to be an SB3 system in the survey of 
\citet{nord1997}. Analysis of 13 spectra showed two measurable components 
with $v_1 \sin i$ = 80 km~s$^{-1}$ and $v_2 \sin i$ = 10 km~s$^{-1}$, 
temperatures $T_1 = 7000 K$, and $T_2 = 6750 K$, and a luminosity ratio of 
0.10. A preliminary orbit for the fainter component was given: $P$ = 5.8167(9)
days, $V_0$ = 11.05(91) km~s$^{-1}$, $K_1$ = 28.9(12) km~s$^{-1}$, 
$e$ = 0.083(56), and $\omega$ = 322(28)\degr. The spectroscopic 
orbit most likely corresponds to the motion of the blend of a faint pair seen
as component B. 

The system was used as a guide star in the MOST 2007 observations of M67. The 
photometry covers three consecutive eclipses. The eclipse depth is very shallow
(only about 0.013 mag) and the eclipses are total (see Fig.~\ref{fig04}). It is
very probable that it is the fainter component of the close visual pair 
(component B) which is the eclipsing binary with its light 
being diluted by the brighter component. In such a 
case, the original, undiluted amplitude of eclipses would be 0.09 mag. Since the 
eclipses are total, the ratio of radii can be determined to be $R_2/R_1 = 
0.3$. The preliminary mid-eclipse ephemeris is HJD $2\,454\,145.85 + 
5.81 \times E$ days.

Our spectroscopy included the AB pair, which could not be separated at the 
spectrograph slit because the typical seeing at the site ($\sim$2 arcsec) is larger 
than the separation of the pair. The broadening functions (Fig.~\ref{fig02})
show the brighter, rapidly-rotating component ($v \sin i \approx 85$ 
km~s$^{-1}$) to be stationary in radial velocity; this star is most likely 
the A component of the visual pair. The fainter component shows radial velocity
variations and is definitely responsible for the eclipses. Our three spectra 
show that the features of the eclipsing pair are always blended, suggesting 
that high-resolution spectroscopy will  be necessary to reliably define orbits
of both components.

\section{$\delta$~Scuti pulsating stars in M67}
\label{deltascuti}

The quality of the MOST data varied over the two runs 
and depended on the position of the target
on the CCD chip mainly because of time and spatial dependence of the stray 
light and/or background. For faint pulsating stars observed
in the M67 field, the instrumentally introduced periodicities affected
the period analysis and its conclusions and limited our ability
to detect very small amplitude pulsations. The direct imaging objects
could in general be followed through the whole satellite orbit so that
the data were entirely suitable for the period analysis but
for the guide stars, 
the background variation led to exclusion of about 20\% of the satellite
period. For both sets of data, however, 
intermittent gaps occurred hence we used the method 
of \citet{deem1975} of period finding with Fourier transform of unevenly spaced
data points. It turned out that all stars observed in the guide star mode
were unsuitable for a sound periodic analysis. Thus, we cannot say anything
about any of the remaining observed blue stragglers, Sanders 752, 968, 977, 
1066, and 1466, which show the frequency spectrum 
dominated by the satellite orbital frequency 14.1994 c/day 
and its integer multiples. The second strong 
periodicity in the data (1 c/day) arises from the periodic 
returns of the satellite to the Earthshine illumination
conditions, which determine the background level. Thus, 
for these stars, any intrinsic variability must have amplitudes smaller
than about 0.001 mag; this limit also applies to Sanders 1263 
which was observed in the raster mode.

Because the direct imaging observations were almost continuous, the
spectral window for such observations is virtually free 
of aliasing. The spectral window for the guide 
stars shows strong side-lobes corresponding the orbital frequency of the satellite. 

\subsection{EX~Cnc}

EX~Cnc (S1284, MMJ~6504) is very probably member of M67 \citep{bala2007}.
Its variability was discovered by \citet{gill1991} and later studied by
\citet{gilbr1992} and \citet{zhzhli2005}. The most detailed investigation of this
$\delta$~Sct variable resulted from the large multi-site campaign of 
\citet{bruntt2007}. The authors presented photometric data from eight sites
using nine telescopes of 0.6 m to 2.1 m, performed over 43 days. 
The authors were able to detect 26 frequencies in their data, but
the spectral window was poor and contained rather strong side-lobes.

\begin{figure}
\includegraphics[width=84mm,clip=]{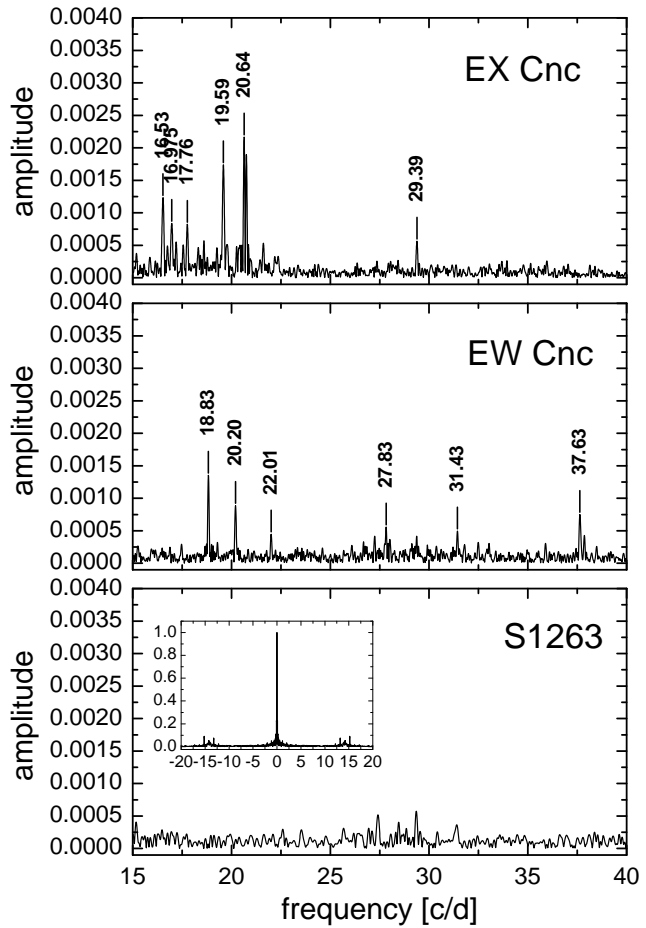}
\caption{The amplitude spectra obtained from the 2007 observations for three 
blue stragglers observed in the raster mode, EX~Cnc, EW~Cnc, 
         and S1263. While the first two objects clearly show the presence of 
         $\delta$~Sct-type pulsations, S1263 does not show any detectable 
         variability in the analyzed frequency range (2 -- 200 c/day). The 
         spectral window given in the S1263 panel is approximately valid 
         for all three stars.
\label{fig07}}
\end{figure}

EX~Cnc was observed by the MOST satellite almost continuously over 30 days
in 2007. The pulsations and their beating are directly visible in the 
data (see Fig.~\ref{fig05}). After the change of the satellite roll angle 
during the 2007 run, the data were of lower quality. For that reason,
only the best and the least interrupted continuous part of 
the run was selected for the period analysis within 
$HJD 2454149.7 - 2454163.1$. 
Although the per-point error is about 0.01 mag, the run is homogeneous and 
the spectral window is practically single peaked; the only side-lobes 
with less than 10\% strength correspond to the satellite orbital cycle 
at the frequencies $\pm$14.2 c/day.  
The amplitude spectrum and the 
corresponding spectral window are shown in Fig.~\ref{fig07}.

The total number of reliably detected frequencies for EX~Cnc is over 20 with
most corresponding to 26 frequencies detected by \citet{bruntt2007}. The accord between 
the results is surprisingly good: out of the 22 strongest frequencies detected by 
\citet{bruntt2007} we could find 20. Because our useful observing window 
is shorter, we could not resolve several close pairs of frequencies resolved 
by the authors. Except for the frequencies found by the authors, we found few 
additional modes at 20.855 c/day, 29.385 c/day, 13.205 c/day and 15.19 c/d. The 
last two, however, could be connected with the orbital period of the satellite.
EX~Cnc was not observed during the 2004 run. All identified frequencies in the amplitude 
spectrum are listed in Table~\ref{tab05}.

\begin{table}
\begin{scriptsize}
\caption{Pulsation frequencies and their amplitudes detected in 
         the 2007 MOST photometry. Frequencies are given both
         in cycles per day and $\mu$Hz; amplitudes are given
         in millimagnitudes. Identifications of particular frequencies
         in \citet{bruntt2007} are given in the last column \label{tab05}
         }
\begin{center}
\begin{tabular}{ccccl}
\hline
No.    & frequency    & frequency &  Amplitude & Identification\\
       & [c/day]      & [$\mu$Hz] &  [mmag]    &       \\
\hline 
\multicolumn{5}{l}{EX~Cnc} \\
 1     & 20.640       & 238.89    &  2.17      & f$_2$ \\
 2     & 20.753       & 240.20    &  1.90      & f$_1$ \\
 3     & 19.589       & 226.72    &  1.74      & $f_{3,5}$  \\
 4     & 16.531       & 191.33    &  1.23      & $f_{4,10}$ \\
 5     & 16.974       & 196.46    &  0.84      & $f_{9,13}$ \\
 6     & 17.761       & 205.57    &  0.82      & $f_6$ \\
 7     & 18.603       & 215.31    &  0.57      & $f_8$ \\
 8     & 29.385       & 340.10    &  0.56      &       \\
 9     & 17.198       & 199.05    &  0.55      & $f_{20}$?  \\
10     & 21.613       & 250.15    &  0.53      & $f_{23}$?  \\
11     & 20.856       & 241.39    &  0.51      &       \\
12     & 17.556       & 203.19    &  0.51      & $f_{17}$   \\
13     & 19.777       & 228.90    &  0.51      & $f_{7,22}$ \\
14     & 20.428       & 236.44    &  0.50      & $f_{16}$   \\
15     & 20.530       & 237.62    &  0.50      &       \\
16     & 16.758       & 193.96    &  0.49      & $f_{14}$   \\
17     & 20.260       & 234.50    &  0.48      & $f_{21}$   \\
18     & 18.321       & 212.05    &  0.46      & $f_{12}$   \\
19     & 19.262       & 222.94    &  0.46      & $f_{18}$   \\
20     & 13.21        & 152.9     &  0.41      &       \\
21     & 15.19        & 175.8     &  0.37      &       \\
22     & 18.42        & 213.2     &  0.35      &       \\
23     & 22.19        & 256.8     &  0.33      & $f_{24}$   \\
24     & 22.36        & 258.8     &  0.33      & $f_{19}$   \\
25     & 18.78        & 217.4     &  0.32      & $f_{11}$   \\
26     & 15.87        & 183.7     &  0.32      &       \\[1.5ex]
\multicolumn{5}{l}{EW~Cnc} \\
1      & 18.834       & 217.99    &  1.35      & $f_1$      \\
2      & 20.198       & 233.77    &  0.89      & $f_2$      \\
3      & 37.632       & 435.56    &  0.75      & $f_5$      \\
4      & 27.828       & 322.08    &  0.56      & $f_4$      \\
5      & 31.432       & 363.80    &  0.50      & $f_3$      \\
6      & 22.008       & 254.72    &  0.49      & $f_6$      \\
7      & 37.86        & 438.2     &  0.43      &            \\
8      & 27.25        & 315.4     &  0.42      & $f_{40}$?  \\
\hline
\end{tabular}
\end{center}
\end{scriptsize}
\end{table}

\subsection{EW~Cnc}

EW~Cnc (S1280, MMJ~5904) is very probably a member of M67 \citep{bala2007}.
The system has a very similar history to EX~Cnc since its discovery by 
\citet{gill1991}, who found that it is a $\delta$~Sct variable. \citet{bruntt2007} 
claimed to detect as many as 41 frequencies in their photometry. Unfortunately, 
EW~Cnc was a rather faint target for the MOST satellite which resulted in a 
0.02 mag point-to-point scatter in the (generally better of the two runs) 
2007 data. The scatter of data 
substantially varied, hence we selected the best and most continuous segment 
for the period analysis (between $HJD 2454145.22 - 2464160$). The 
brightness variations caused by the $\delta$~Sct-type pulsations are barely 
detectable by visual inspection of the light curve. Nevertheless, eight 
frequencies are clearly visible in the 2007 amplitude spectrum; the strongest six
practically coincide with those detected in the more precise data of both
\citet{gilbr1992} and \citet{bruntt2007}. EW~Cnc was also observed during 
the shorter observing run in 2004. The amplitude spectrum shows three frequencies:
18.83, 27.85 and 20.20 c/day; the same as detected from the 2007 run data 
(see Fig.~\ref{fig07}, Table~\ref{tab05}) and found by \citet{bruntt2007}. 
This indicates that the pulsation modes are highly stable over long time intervals.

\section{Conclusions}
\label{sum}

Two long, almost continuous photometric runs (10 days in 2004 and 30 days in 2007) of
stars in the M67 field were obtained by the MOST space mission, followed by 
ground based spectroscopic observations at DDO. The targets include eclipsing 
blue stragglers in M67, $\delta$~Scuti-type pulsating star, and variable
non-cluster members in the cluster field.

Among the main results, our analysis of the light curve of the contact binary 
AH~Cnc supports its low mass ratio and a high inclination angle, as found by 
\citet{sand2003a}. Contrary to a recent suggestion, there is no compelling 
evidence for a third component in the system. The MOST light curves of the 
blue straggler triple system ES~Cnc show a clear asymmetry which we interpret 
as the presence of two dark photospheric spots, about 1200~K cooler than the 
surrounding photosphere. A comparison of the 2004 and 2007 photometry reveals 
almost no changes in the fitted sizes and positions of the spots, suggesting 
a surprising stability in the photospheric features of the components over 
three years. However, our analysis of ES~Cnc is only preliminary. The system 
deserves a dedicated campaign of high-resolution spectroscopy and simultaneous 
high-precision photometry, but the proximity of its period to one day is a 
somewhat complicating factor.

Two new field eclipsing systems were discovered: HD~75638 and GSC~814-323. 
The former is a known spectroscopic triple. The eclipses are very shallow 
($A \sim 0.013$ mag) and would be barely detectable from the ground. The 
photometric period (the mean interval between the eclipses) of 5.81 days is 
compatible with the spectroscopic period of 5.8167 days found from observations 
spanning 3.68 years by \citet{nord1997}. High-resolution spectroscopy of the 
system combined with our light curve could lead to reliable determination of 
masses of all three components. Spectra of GSC~814-323 show that it is a 
spectroscopic binary. Since the ratio of intensities of the components in the 
broadening functions is about 0.29, we expect the secondary eclipses to be 0.09 
mag deep.

Period analysis of two known $\delta$~Scuti variables, EX~Cnc and EW~Cnc 
yielded a very reliable frequency spectrum without any obvious instrumental
periodicities. For the brighter of the two variables, EX~Cnc, we could 
reliably determine as many as 26 pulsational frequencies. In spite of the
larger photometric scatter of about 0.02 mag for the fainter variable, 
EW~Cnc, we see 8 pulsation frequencies in the amplitude spectrum. Pulsational frequencies 
in both stars are temporary highly stable. The guide star photometry of other 
blue stragglers (Sanders 752, 968, 977, 1066, and 1466) 
was affected by the strong background variations which limited amplitudes
of any detectable pulsations to $<0.001$ mag. 

\section*{Acknowledgments}
This study has been funded by the Canadian Space Agency Space Enhancement 
Program (SSEP) with TP holding a Post-Doctoral Fellowship position at the 
University of Toronto. The Natural Sciences and Engineering Research Council 
of Canada (NSERC) supports the research of DBG, JMM, AFJM, and SMR. Additional
support for AFJM comes from FQRNT (Quebec). RK is supported by the Canadian 
Space Agency and WWW is supported by the Austrian Space Agency and the Austrian 
Science Fund (P17580). This research has made use of the SIMBAD database, 
operated at CDS, Strasbourg, France and NASA's Astrophysics Data System 
Bibliographic Services.

\label{lastpage}
\end{document}